\documentclass[twocolumn,english,prl,showpacs,floatfix]{revtex4}
\usepackage{amsmath}
\usepackage{amssymb}
\usepackage{graphicx}
\usepackage{color}
\makeatletter
\@ifundefined{textcolor}{}
{%
 \definecolor{BLACK}{gray}{0}
 \definecolor{WHITE}{gray}{1}
 \definecolor{RED}{rgb}{1,0,0}
 \definecolor{GREEN}{rgb}{0,1,0}
 \definecolor{BLUE}{rgb}{0,0,1}
 \definecolor{CYAN}{cmyk}{1,0,0,0}
 \definecolor{MAGENTA}{cmyk}{0,1,0,0}
 \definecolor{YELLOW}{cmyk}{0,0,1,0}
 }

\@ifundefined{showcaptionsetup}{}{%
 \PassOptionsToPackage{caption=false}{subfig}}
\makeatother

\usepackage{babel}

\begin{document}

\title{Heat Transport between Antiferromagnetic Insulators and Normal Metals}

\begin{abstract}
Antiferromagnetic insulators can become active spintronics components by controlling and detecting their dynamics via spin currents in adjacent metals. This cross-talk occurs via spin-transfer and spin-pumping, phenomena that have been predicted to be as strong in antiferromagnets as in ferromagnets. Here, we demonstrate that a temperature gradient drives a significant heat flow from magnons in antiferromagnetic insulators to electrons in adjacent normal metals. The same coefficients as in the spin-transfer and spin-pumping processes also determine the thermal conductance. However, in contrast to ferromagnets, the heat is not transferred via a spin Seebeck effect which is absent in antiferromagnetic insulator-normal metal systems. Instead, the heat is transferred via a large staggered spin Seebeck effect.
\end{abstract}

\author{Arne Brataas}
\email{Arne.Brataas@ntnu.no}
\affiliation{Department of Physics, Norwegian University of Science and Technology, NO-7491 Trondheim, Norway}

\author{Hans Skarsv{\aa}g}
\affiliation{Department of Physics, Norwegian University of Science and
Technology, NO-7491 Trondheim, Norway}

\author{Erlend G. Tveten}
\affiliation{Department of Physics, Norwegian University of Science and
Technology, NO-7491 Trondheim, Norway}

\author{Eirik L\o haugen Fj\ae rbu}
\affiliation{Department of Physics, Norwegian University of Science and
Technology, NO-7491 Trondheim, Norway}

\pacs{72.25.Mk,72.20.Pa,73.50.Lw,72.10.Di}
\maketitle

In spintronics, the properties which make antiferromagnets markedly different from ferromagnets also make them attractive in a more dynamic role. Antiferromagnets operate at much higher frequencies and may empower Terahertz circuits. They also have no magnetic stray fields, which therefore enables denser spintronics circuits. For these reasons, antiferromagnets are usually passive spintronics components. However, they can play a role as active components despite their lack of a macroscopic magnetic moment \cite{Nunez:prb2006,Wei:prl2007,Urazhdin:prl2007,Haney:prl2008,Xi:prl2008,Gomonay:ltphys2008,Hals:prl2011,Park:nmat2011,Marti:prl2012,Tveten:prl2013,Marti:nmat2014,Tveten:prl2014,Cheng:prl2014} and even when they are insulating \cite{Tveten:prl2013,Tveten:prl2014,Cheng:prl2014}. 

We demonstrate that the thermal coupling between antiferromagnetic insulators (AFIs) and normal metals is relatively strong. The strong thermal coupling facilitates several outcomes, can lead to efficient cooling of antiferromagnetic spintronics devices, might function as heat sensors and can reveal valuable information about the high-frequency spin excitations in DC measurements that are complicated to extract with other techniques.

Antiferromagnets can produce pure spin currents as large as those produced by ferromagnets. We recently showed that spin pumping may be as operative from antiferromagnets as from ferromagnets  \cite{Cheng:prl2014}, in apparent contraction to naive intuition.   Furthermore, the efficiency of spin pumping from antiferromagnets to normal metals implies, via Onsager reciprocity relations, that there is a considerable spin-transfer torque on antiferromagnets from a spin accumulation in adjacent normal metals.  However, in the absence of external magnetic fields, the spin Seebeck effect vanishes \cite{Ohnuma:prb2013}. This fact seems to indicate that spins in antiferromagnets decouple from, or are only weakly connected to, heat currents and temperature gradients in adjacent normal metals.

To the contrary, we find that the thermal coupling constant is orders of magnitude stronger than its ferromagnetic counterpart. 
This radical difference is caused by the large exchange field in antiferromagnets that governs the heat transfer rather than the much smaller anisotropy fields or external magnetic fields in ferromagnets. The thermal coupling between antiferromagnetic insulators and normal metals is associated with a staggered spin Seebeck effect rather than via the spin Seebeck effect. 

Spin caloritronics determines how  spins are coupled to currents and temperature gradients \cite{Bauer:nmat2012}. Measurements of important thermoelectric properties in ferromagnetic insulators, such as the spin Seebeck effect \cite{Uchida:nat2008}, are central to this field. In the spin Seebeck effect, a temperature gradient transfers a magnon spin current in a ferromagnet into an itinerant spin current in a normal metal \cite{Xiao:prb2010,Adachi:prb2011}. This process is active even in insulating ferromagnets \cite{Uchida:nmat2010}. The spin Peltier effect is reciprocal to the spin Seebeck effect; a heat current generates a spin accumulation \cite{Flipse:nntech2012,Flipse:prl2014}. These fascinating thermoelectric properties can be useful to control the heat flow in spintronics devices and in devices that recycle waste heat. 

In explaining our calculations, we interpret the theories on the spin Seebeck effect \cite{Uchida:nat2008,Xiao:prb2010,Adachi:prb2011,Bauer:nmat2012,Hoffman:prb2013} as a combination of three mechanisms. First, a precessing magnetization can pump a spin current across a ferromagnet-normal metal junction \cite{Tserkovnyak:prl2002,Brataas:prb2002,Tserkovnyak:rmp2005}. Spin-pumping gives rise to an increased magnetization dissipation rate \cite{Mizukami:jjap2001,Urban:prl2001,Tserkovnyak:prl2002}. Second, the enhanced dissipation implies that there is also an enhanced spin current noise in terms of a fluctuating spin-transfer torque \cite{Foros:prl2005}. In equilibrium,  there is no thermal bias and the DC spin current vanishes because the temperature-driven spin pumping and a fluctuating spin-transfer torque exactly compensate each other. Third, a temperature difference alters this balance and causes a net spin current \cite{Xiao:prb2010,Adachi:prb2011,Hoffman:prb2013}. 

In this picture, to compute the heat transfer between AFIs and normal metals, we first establish the fluctuating spin-transfer and staggered spin-transfer  in such hybrid systems. Both quantum {\it and} thermal fluctuations are required to determine the magnon occupations. Subsequently, we use these results to define the thermal  gradient driven (staggered) spin currents, which we then use to evaluate the rate of change of the heat in the AFI. We focus on insulating antiferromagnets where the transport properties are magnon-driven. Generalizations to conducting antiferromagnets are straightforward.   

We model the AFI as a two sub-lattice system with spatio-temporal magnetizations $\mathbf{M}_1$ and $\mathbf{M}_2$. The dynamics are described by the staggered magnetizations $\mathbf{L}=\mathbf{M}_1-\mathbf{M}_2=L \mathbf{n}$ and the magnetization $\mathbf{M}=\mathbf{M}_1+\mathbf{M}_2=L\mathbf{m}$. These fields satisfy the constraints $\mathbf{n}^2+\mathbf{m}^2=1$ and $\mathbf{n} \cdot \mathbf{m}=0$. In equilibrium, the staggered field is homogeneous and constant in time, $|\mathbf{L}|=L$, and the magnetization vanishes, $\mathbf{M}=0$, i.e. $\mathbf{n}^2=1$ and $\mathbf{m}=0$. We consider an easy-axis AFI that is described by the free energy $F=\int d\mathbf{r} \left[ f(\mathbf{r}) + f_s(\mathbf{r})\right]$, where the energy density is 
\begin{equation}
f = \frac{L}{\gamma} \left[\frac{1}{2} \omega_E (\mathbf{m}^2-\mathbf{n}^2) - \frac{1}{2} \omega_A (m_z^2 + n_z^2)  \right] , \label{eq:f_endens}
\end{equation}
with the exchange frequency $\omega_E$ and the smaller anisotropy frequency $\omega_A$, $\omega_A \ll \omega_E$. When $\mathbf{n}$ and $\mathbf{m}$ spatially vary, the stiffness contributions are
\begin{equation}
f_s = \frac{L}{2\gamma} \omega_A \sum_{i=x,y,z}\left[(\lambda_n \partial_i \mathbf{n})^2 +  (\lambda_m \partial_i \mathbf{m})^2 \right] \, , \label{eq:f_enstiff}
\end{equation}
where $\lambda_n$ and $\lambda_m$ are exchange lengths associated with $\mathbf{n}$ and $\mathbf{m}$, respectively. The dynamic equations are:
\begin{subequations}
\begin{align}
\dot{\mathbf{n}} &= {\boldsymbol{\omega}}_{m} \times \mathbf{n} + {\boldsymbol{\omega}}_{n} \times \mathbf{m} + \boldsymbol{\tau}_n \, , \label{eq:eomn}\\
\dot{\mathbf{m}} &= {\boldsymbol{\omega}}_{n} \times \mathbf{n} + \boldsymbol{\omega}_{m} \times \mathbf{m}+ \boldsymbol{\tau}_m \, , \label{eq:eomm}
\end{align}
\label{eq:eom}
\end{subequations}
where the effective fields are $\boldsymbol{\omega}_n= -(\gamma/L) \delta f/\delta  \mathbf{n}$ and $\boldsymbol{\omega}_m= -(\gamma/L) \delta f/\delta  \mathbf{m}$. In  Eqs.\ (\ref{eq:eomn}) and (\ref{eq:eomm}), the dissipation and fluctuation torques $\boldsymbol{\tau}_m$  and $\boldsymbol{\tau}_n$ are essential to describe spin caloritronics effects.  
\begin{figure}
\centering
\includegraphics[scale=0.45]{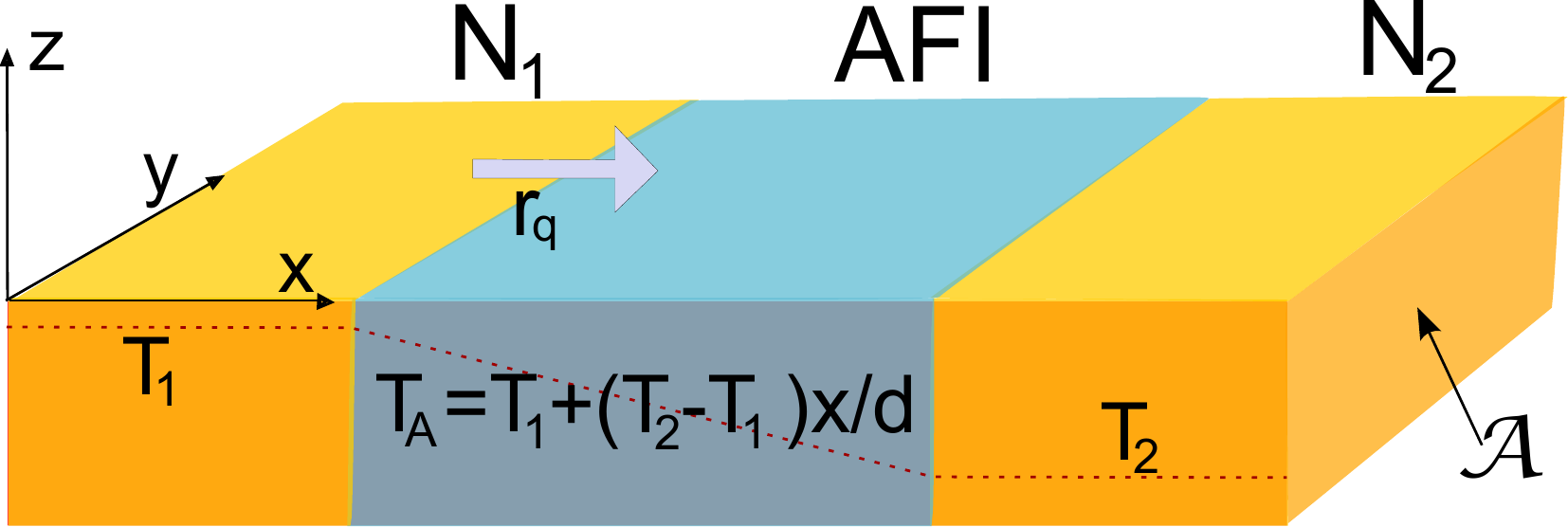}
\caption{An antiferromagnet insulator (AFI) sandwiched between two normal metals $N_1$ and $N_2$. The left normal metal ($N_1$) is a good spin sink. The right normal metal ($N_2$) has no spin memory loss. A heat current between the normal metals flows in response to an applied temperature gradient across the AFI. The cross section is ${\cal A}$ and $d$ is the AFI thickness. The heat flow $I_Q$ is along the longitudinal coordinate $x$.}
\label{fig:system}
\end{figure}

We consider a thin-film AFI of thickness $d$ sandwiched between two normal metals, the left one of which is a good spin sink (e.g.\ Pt), and the right one has little or no spin memory loss (e.g. Cu), see Fig.\ \ref{fig:system}. We assume planar AFI-normal metal interfaces of cross section ${\cal A}$. The coordinate ${\bf r}=(x,\boldsymbol{\rho})$ is decomposed into a perpendicular coordinate $x$ ($0\le x \le d$) and the 2D in-plane coordinates $\boldsymbol{\rho}$.  The fluctuation-dissipation torques have bulk and (spin-pumping induced) interface contributions, $\tau_\nu({\bf r})=\tau^{(b)}_\nu({\bf r})+\delta(x-x_I)\tau^{(p)}_\nu(\boldsymbol{\rho})$, where $x_I=0^+$ is infinitesimally near the interface on the AFI side close to the spin sink and the subscript $\nu$ denotes the product of either the sub-index $n$ or $m$ and a Cartesian component $x$, $y$, or $z$.  The bulk torques arise from the magnon-phonon interaction. At the AFI-N interface, the torques are governed by spin-pumping induced by the coupling of the magnetic moments to itinerant electrons in adjacent normal metals.  In finding the torques, we introduce dissipation in a similar way as in Ref.\ \onlinecite{Hals:prl2011} and further generalize this description to include quantum and thermal fluctuations. The resulting fluctuation-dissipation torques are
\begin{subequations}
\begin{align}
\boldsymbol{\tau}_n &= (\mathbf{h}_{m}- \alpha \dot{\mathbf{m}}) \times \mathbf{n} + (\mathbf{h}_n- \alpha \dot{\mathbf{n}}) \times \mathbf{m}\, ,  \\
\boldsymbol{\tau}_m & = (\mathbf{h}_n - \alpha \dot{\mathbf{n}}) \times \mathbf{n} + (\mathbf{h}_m -\alpha \dot{\mathbf{m}}) \times \mathbf{m} \, ,
\end{align} 
\end{subequations}
for both bulk and interface contributions where we have suppressed the superscript [$(b)$ or $(p)$] in $\boldsymbol{\tau}$, $\mathbf{h}$, and $\alpha$. The bulk Gilbert damping constant is $\alpha^{(b)}$. $\alpha^{(p)}$ is a measure of the spin-pumping-induced enhanced dissipation; for homogenous macrospin excitations  the enhanced damping constant is $\alpha^{(b)}+\alpha^{(p)}/d$ \cite{Tserkovnyak:rmp2005}. 

The fluctuation-dissipation theorem implies the existence of the fluctuating forces $\mathbf{h}_m$ and $\mathbf{h}_n$. The average of the independent fluctuating forces $\mathbf{h}_m$ and $\mathbf{h}_n$ as well as all their fluctuations vanish, except
\begin{subequations}
\begin{align}
\langle h_{\nu }^{(p)}(\boldsymbol{\rho}t) h_{\nu}^{(p)} (\boldsymbol{\rho}'t') \rangle & = \frac{\gamma \alpha^{(p)} R(t \! - \! t',T_1)}{L \pi} \delta(\boldsymbol{\rho} \! - \!  \boldsymbol{\rho}')   
\label{eq:fluctpumping}
 \, , \\
\langle h_{\nu}^{(b)}({\bf r}t) h_{\nu}^{(b)} ({\bf r}'t') \rangle &= \frac{\gamma \alpha^{(b)} R(t \! - \! t',T_A)}{L \pi} \delta({\bf r} \! - \!  {\bf r}')    \, .
\label{eq:fluctbulk}
\end{align}
\end{subequations}
The correlation function $R(t,T)$ depends on the (local) temperature.   As demonstrated for ferromagnets in Ref.\ \onlinecite{Foros:prl2005}, the spin-current fluctuations associated with spin-pumping depend on the temperature in the normal metal close to the interface, $T_1$. We posit that the one-to-one correspondence between spin-pumping in ferromagnets and antiferromagnets \cite{Cheng:prl2014} implies that the spin-current fluctuations in antiferromagnets obeys the same relationship, as in Eq.\ (\ref{eq:fluctpumping}). In the bulk of the AFI, the phonon-induced fluctuations associated with the bulk Gilbert damping depends on the temperature profile in the antiferromagnet $T_A(x)$, as in Eq.\ (\ref{eq:fluctbulk}). The correlation function only describes white noise in the (classical) high-temperature limit, $R(t,T) \approx 2 \pi k_B T \delta(t)$. However, for our purposes, the magnon occupations also depend on the low-temperature quantum limit of the fluctuations, which we describe after Eq.\ \eqref{eq:rqdos}.

The effective fields determined by Eqs. \eqref{eq:f_endens} and \eqref{eq:f_enstiff} are
\begin{subequations}
\begin{align}
\boldsymbol{\omega}_n &=   \omega_E \mathbf{n} +  \omega_A (\mathbf{n} \cdot \hat{z}) \hat{z} + \omega_A (\lambda_n \nabla)^2 \mathbf{n} \label{omegan} \, , \\
\boldsymbol{\omega}_m & = -\omega_E \mathbf{m} +  \omega_A (\mathbf{m} \cdot \hat{z}) \hat{z}+  \omega_A (\lambda_m \nabla)^2 \mathbf{m} 
\label{omegam}  \, .
\end{align}
\end{subequations}
We focus on the rate of change of the heat ($I_Q$) from the AFI into the normal metal. Evaluating the rate of change of the free energy  due to the (spin-pumping) interface fluctuation-dissipation torques, $I_Q=L \gamma^{-1} \int d\boldsymbol{\rho} [\boldsymbol{\omega}_n \cdot \boldsymbol{\tau}^{(p)}_{n}+ \boldsymbol{\omega}_m \cdot \boldsymbol{\tau}^{(p)}_{m}  ]$ and using the continuity equations for the spin and staggered spin currents below in Eqs.\ (\ref{eq:linbound}), we find
\begin{equation}
I_Q  = -\frac{L {\cal A} d}{\gamma}\text{Im} \int \frac{d \boldsymbol{\rho}}{{\cal A}} \left[ m_+ \omega_{E}\omega_A \lambda_n^2 \frac{\partial n_-}{d \partial x} \right] \, , 
\end{equation}
where we use a circular basis so that $n_\pm=n_x \pm i n_y$ and $m_\pm = m_x \pm i m_y$ are first order corrections with respect to the equilibrium configuration $\mathbf{n}=\hat{z}$ and $\mathbf{m}=0$. $I_Q$ has units of energy per time.

Next, we Fourier transform in the transverse coordinate $\boldsymbol{\rho}$ and time $t$ so that any function $c(x,\boldsymbol{\rho},t)=\sum_{\mathbf{q}} \int d\omega \tilde{c}(x,\mathbf{q},\omega)\exp{i (\omega t - \mathbf{q} \cdot \boldsymbol{\rho})}$. Using Eq.\ (\ref{eq:eom}), the linearized dynamic equations of motion become
\begin{widetext}
\begin{subequations}
\begin{align}
\left\{ i \alpha^{(b)} \omega + \omega_A \left[1 + \lambda_m^2 (\mathbf{q}^2 - \partial_x^2)\right] +  2 \omega_E \right\} \tilde{m}_\pm  = \pm \omega \tilde{n}_\pm + \tilde{h}_{m\pm}^{(b)}(x) \, ,
\label{eq:lineomm}\\
\left\{ i \alpha^{(b)} \omega + \omega_A \left[(1 + \lambda_n^2 (\mathbf{q}^2 - \partial_x^2)\right]  \right\} \tilde{n}_\pm  = \pm \omega \tilde{m}_\pm + \tilde{h}_{n\pm}^{(b)}(x) \, .
\label{eq:lineomn}
\end{align}
\label{eq:lineom}
\end{subequations}
\end{widetext}
In the coupled dynamic equations (\ref{eq:lineom}), the stiffness contributions \eqref{eq:f_enstiff} can be interpreted as arising from the continuity equations for the staggered field and the magnetizations, $(\partial_t \mathbf{n})_s+ \sum_i \partial_i \mathbf{j}_{n,i}=0$ and $(\partial_t \mathbf{m})_s + \sum_i \partial_ i \mathbf{j}_{m,i}=0$. In linear response, the staggered spin current and spin currents along the $x$-direction  are $\mathbf{j}_{n,x}= \omega_A \lambda_m^2 \hat{z} \times \partial_x \mathbf{m}$ and $\mathbf{j}_{m,x}= \omega_A \lambda_n^2 \hat{z} \times \partial_x \mathbf{n}$. The boundary conditions for the linearized equation of motion (\ref{eq:lineom}) are obtained by integrating the dynamic equations (\ref{eq:eom}) across the AFI-N interface. This results in the continuity of the spin and staggered spin currents in linearized forms  at $x=0$:
\begin{subequations}
\begin{align}
\omega_A \lambda_m^2 \frac{\partial \tilde{m}_\pm}{\partial x}  & = i \omega \alpha^{(p)}  \tilde{m}_\pm- \tilde{h}_{m,\pm}^{(p)} \, ,
\label{eq:linboundm}\\
\omega_A \lambda_n^2 \frac{\partial \tilde{n}_\pm}{\partial x}  & = i \omega \alpha^{(p)} \tilde{n}_\pm  - \tilde{h}_{n,\pm}^{(p)} \, .
\label{eq:linboundn}
\end{align}
\label{eq:linbound}
\end{subequations}
Similarly, at $x=d$, there is no loss of currents and the boundary conditions are $\omega_A \lambda_m^2 \partial \tilde{m}_\pm / \partial x=0$ and $\omega_A \lambda_n^2 \partial \tilde{n}_\pm / \partial x=0$.

In typical antiferromagnets, $\omega_E$ is much larger than all other energy scales and we may employ the so-called exchange approximation. This implies that we may disregard smaller terms in the equation of motion (\ref{eq:lineomm}) so that it greatly simplifies to $\tilde{m}_\pm = \omega \tilde{n}_\pm/2\omega_E$. By inserting this relation into Eq.\ (\ref{eq:lineomn}), we find the equation of motion in the exchange approximation
\begin{equation}
\lambda_n^2 (q_x^2 + \partial_x^2) \tilde{n}_\pm = -\frac{\tilde{h}^{(b)}_{n\pm}}{\omega_A} \, 
\label{eq:closedeom}
\end{equation}
which can be solved with the boundary conditions of Eq.\ (\ref{eq:linboundn}). In the exchange approximation, to the lowest order in the dissipation, we have introduced the longitudinal wavenumber $q_x$. The complex wavenumber $q_x$ is implicitly defined via the relation $\omega= \omega_R + i/t^{(b)}$, where the bulk resonance frequency and the bulk lifetime are determined by
\begin{subequations}
\begin{align}
\omega_R^2 & = 2 \omega_A \omega_E (1 + \lambda_n^2 (q_x^2 + q_y^2+q_z^2))  \, , 
\label{eq:omegaR}\\
1/t^{(b)} & = \alpha^{(b)} \omega_E  \, .
\label{eq:lifetimebulk}
\end{align}
\end{subequations}
The central results we will obtain can be interpreted in terms of the eigenstates with the associated eigenfrequencies and lifetimes in a thin film antiferromagnet. The eigenstates are determined by expressing  $\tilde{n}_\pm=  A_\pm \exp{(i q_x x)} + B_\pm \exp{(- i q_x x)}$ in Eq.\ (\ref{eq:closedeom}) when the right hand side (the fluctuations) vanishes. The only non-trivial solution that satisfies both the boundary conditions of Eqs.\ (\ref{eq:linboundn}) at $x=0$ (with no fluctuations) and $\partial \tilde{n}_\pm/\partial x=0$ at $x=d$ is determined by the secular equation $s(q_x)=0$, where
\begin{equation}
s(q_x) = \frac{q_x \lambda_n^2 \omega_A}{ d\omega } \tan{(q_x d)} - i \frac{\alpha^{(p)}}{d} \, .
\label{eq:secular}
\end{equation}
In the absence of spin-pumping and bulk damping, the solutions of $s(q_x)=0$ are standing waves where $q_x=N \pi/d$ and $N$ is an integral number. When spin-pumping is weak, the second term in Eq.\ (\ref{eq:secular}) is small and the solutions of $s({q_x})=0$ can be expanded around the solutions obtained in the absence of spin-pumping. For the higher modes, when $N \ne 0$, we expand the wavevector $q_x$ to the first order in the deviations from $N \pi/d$ and insert the resulting imaginary part of the wavevector  into the dispersion relation of Eq.\ (\ref{eq:omegaR}) to find the spin-pumping lifetime $t_N^{(p)}$. For $N=0$, we carry out a second-order expansion in terms of the small parameter $q_x d$ around 0 and insert this result into the dispersion of Eq.\ (\ref{eq:omegaR}) to find the lifetime $t_0^{(p)}$. We compute that 
\begin{subequations}
\begin{align}
	1/t_0^{(p)} = \frac{\alpha^{(p)}}{d} \omega_E \, , \\
	1/t_{N\ne0}^{(p)} = 2  \frac{\alpha^{(p)}}{d} \omega_E \, . 
\end{align}
\label{eq:lifetimepumping}
\end{subequations}
In a striking contrast to ferromagnets, the spin-pumping induced scattering rate $1/t^{(p)}$ \eqref{eq:lifetimepumping} is proportional to the exchange energy rather than the spin-wave eigenfrequency. This is in agreement with the behavior for the bulk lifetime (\ref{eq:lifetimebulk}), and, as we will see, implies a large heat conductance between AFIs and normal metals. Interestingly, we find that the spin-pumping induced relaxation rate of the higher modes is twice as large as the uniform, but independent of the transverse (2D) wavevector $\mathbf{q}$. This ratio agrees with our previous result for the spin-pumping induced ratio in thin film ferromagnets and can be used to distinguish the spin-wave modes \cite{Kapelrud:prl2013}.

Next, we solve the linearized dynamic equation (\ref{eq:lineom}) with the fluctuating bulk forces and subject to the boundary condition (\ref{eq:linboundn}) where the fluctuating spin-pumping induced forces appear. To compute the rate of change of the heat, we represent the solution at $x=0$ as $\tilde{m}_+ =\chi_{m+}^{(p)} \tilde{h}_{n+}^{(p)} + \int_0^d dx \chi_{m+}^{(b)}(x) \tilde{h}_{n+}^{(b)} (x)$ and $\omega_A \lambda_n^2 \partial_x \tilde{n}_+/d =\chi_{n'+}^{(p)} \tilde{h}_{n+}^{(p)} + \int_0^d dx \chi_{n'+}^{(b)}(x) \tilde{h}_{n+}^{(b)} (x) $. We find that $\chi_{m+}^{(p)}=- 1/(2 d \omega_E s(q_x))$, $\chi_{m+}^{(b)}= \chi_{m+}^{(p)} \cos{q_x (d-x)}/\cos{q_x d}$,  $\chi_{n'+}^{(p)}=- q_x \lambda^2 \omega_A \tan{q_x d}/d^2 \omega s(q_x)$, and $\chi_{n'+}^{(b)}= -i\alpha^{(p)} \cos{q_x(d-x)}/d^2 s(q_x) \cos{q_x d}$.

We evaluate the variance of the fluctuating forces and find the rate of change of the heat, $I_Q= (2 d \omega_E/\pi) \text{Im}\sum_\mathbf{q} \int_{-\infty}^\infty d\omega[\eta^{(p)}_{Q}+\eta^{(b)}_{Q}]$, where the spin-pumping and bulk contributions are 
\begin{subequations}
\begin{align}
	\eta_Q^{(p)} & =    \chi_{m+}^{(p)} (\chi_{n'+}^{(p)})^*  \alpha^{(p)} \tilde{R}(\omega,T_1) \, , \\
	\eta_Q^{(b)} & =   \int_0^d dx \chi_{m+}^{(b)}(x) (\chi_{n'+}^{(b)}(x))^*  \alpha^{(b)} \tilde{R}(\omega,T_A(x)) \, .
\end{align}
\end{subequations}
In equilibrium, $T_A(x)=T_1$, the rate of change of the heat vanishes, $I_Q=0$,  as expected. In linear response, the temperature varies linearly in the AFI so that $T_A(x)=T_1 + (T_2-T_1) x/d$. We then compute that the rate of change of the heat is $I_Q=(T_2-T_1) \kappa_Q$, where
\begin{equation}
	\kappa_Q = \sum_{q_y,q_z} \int_{-\infty}^\infty d\omega \frac{\alpha^{(b)} \frac{\alpha^{(p)} }{d}}{2 \pi |s(q_x)|^2} \zeta(q_x) \frac{\partial \tilde{R}(\omega,T)}{\partial T}
	\label{eq:rqgeneral}
\end{equation}
and $\zeta(q_x)=2 \int_0^d dx |\frac{\cos{q_x(d-x)}}{\cos{q_xd}}|^2 \frac{x}{d^2}$. By following the same methods, we also compute that the temperature-driven spin current, i.e. the spin Seebeck effect, vanishes, in agreement with Ref.\ \onlinecite{Ohnuma:prb2013}. However, we find that the temperature-driven staggered spin current is finite. The heat is therefore transferred between the AFI and the normal metal by means of a staggered spin current. 

By comparing the equilibrium expectation value of the spin-wave free energy with the quantum-mechanical result for a magnon gas or, alternatively, by using the fluctuation-dissipation theorem represented by Eq. (4.9) in Ref.\ \onlinecite{Callen:prb1951}, we identify that the correlations function $R(\omega,T)$ represents the mean energy at the temperature $T$ of an oscillator at natural frequency $\omega$, $
\tilde{R}(\omega,T) = \frac{1}{2} \hbar \omega + \hbar \omega f(\omega,T)$, where $f(\omega,T)$ is the Bose-Einstein distribution function. At high temperatures we thus obtain the classical $\tilde{R} \approx k_B T$, as expected. However, the complete frequency dependence of $\tilde{R}(\omega,T)$ is essential in determining the thermally active magnon modes and quantifying the heat conductance $\kappa_Q$.
\begin{widetext}
In the limit of low bulk damping and spin-pumping induced damping, we can expand the poles of the denominator of Eq.\ (\ref{eq:rqgeneral}) around the spin-wave resonance $q_x d= n \pi$ in a similar way as in Ref. \onlinecite{Hoffman:prb2013}. This results in an intuitive expression:
\begin{equation}
	I_Q = \sum_{N=0}^\infty \frac{1}{t_N^{(p)}} \int_0^\infty d\omega D_N(\omega) \hbar \omega \left\{f(\omega,T_1) \left[1-f(\omega,T_2)\right]-f(\omega,T_2)\left[1-f(\omega,T_1)\right]\right\} \, .
	\label{eq:rqdos}
\end{equation}
\end{widetext}
The heat current that flows between the normal metals via the antiferromagnet, at each frequency, is proportional to the spin-pumping induced spin-wave relaxation rate $1/t_N^{(p)}$, the mode-dependent density of states, $D_N(\omega)=\sum_{q_y,q_z} \delta(\omega-\left\{2 \omega_A \omega_E \left[1+\lambda_n^2 ((\frac{N \pi}{d})+q_y^2+q_z^2)^2\right]\right\}^{1/2})$. Furthermore,  the heat current is determined by the Bose-Einstein occupation of the magnons and the electron-hole pairs in the normal metal. This expression (\ref{eq:rqdos}) reveals that the thermal coupling between normal metals and AFIs is relatively strong. The heat current is proportional to the spin-pumping induced spin-wave scattering rates that are proportional to the exchange energy and the Gilbert damping coefficient.  At high temperature, we find $I_Q = {\cal A}  \pi^2 (k_B T_1)^3 k_B (T_{2}-T_{1}) \alpha^{(p)}/(15 \sqrt{2} A_{ex}^{3/2} \sqrt{\omega_E} \hbar^{3/2} )$, where $A_{ex}= \hbar \omega_A \lambda_n^2$ is the exchange stiffness.

Phonons  also mediate heat currents between AFIs and normal metals. Experimentally, the magnon-induced heat current we predict here can be separated from the phonon heat current by the different temperature and length dependence. For instance, different measurements of normal metals with strong and weak spin-orbit coupling can be compared. Additionally, one can use an external magnetic field to change the magnon dispersion and consequently the spin-wave density of states governing magnon-induced heat current of Eq.\ (\ref{eq:rqdos}).

In conclusion, we demonstrated a strong thermal coupling between antiferromagnetic insulators and normal metals. The heat is carried in terms of a staggered spin current. 

We acknowledge support from the Research Council of Norway, project number 216700.


\begin{thebibliography}{99}
	
\bibitem{Nunez:prb2006} A. S. Nunez, R. A. Duine, P. Haney, and A. H. MacDonald, Phys. Rev. B \textbf{73}, 214426 (2006).
	
	
	
\bibitem{Wei:prl2007} Z. Wei, A. Sharma, A. Nunez, P. M. Haney, R. A. Duine, J. Bass, A. H. MacDonald, and M. Tsoi, Phys. Rev. Lett. \textbf{98}, 116603 (2007).
	
\bibitem{Urazhdin:prl2007} S. Urazhdin and N. Anthony, Phys. Rev. Lett. \textbf{99}, 046602 (2007). 
	
\bibitem{Haney:prl2008} P. M. Haney and A. H. MacDonald, Phys. Rev. Lett. \textbf{100}, 196801 (2008).
	
\bibitem{Xi:prl2008} Y. Xu, S. Wang, and K. Xia, Phys. Rev. Lett. \textbf{100}, 226602 (2008).
	
\bibitem{Gomonay:ltphys2008} H. V. Gomonay and V. M. Loktev, Low Temp. Phys. \textbf{34}, 198 (2008). 
	
\bibitem{Hals:prl2011} K. M. D. Hals, Y. Tserkovnyak, and A. Brataas, Phys. Rev. Lett. \textbf{106}, 107206 (2011).
	
\bibitem{Park:nmat2011} B. G. Park, J. Wunderlich, X. Marti, Y. Kurosaki, M. Yamada, H. Yamamoto, A. Nishide, J. Hayakawa, H. Takahashi, A. B. Shick and T. Jungwirth, Nature Mat. \textbf{10}, 347 (2011).
	
\bibitem{Marti:prl2012} X. Marti, B. G. Park, J. Wunderlich, H. Reichlova, Y. Kurosaki, M. Yamada, H. Yamamoto, A. Nishide, J. Hayakawa, H. Takahashi, and T. Jungwirth, Phys. Rev. Lett. \textbf{108}, 017201 (2012).
	
\bibitem{Tveten:prl2013} E. G. Tveten, A. Qaiumzadeh, O. A. Tretiakov, and A. Brataaas, Phys. Rev. Lett. \textbf{110}, 127208 (2013).
	
\bibitem{Marti:nmat2014} X. Marti, I. Fina, C. Frontera, J. Liu, P. Wadley, Q. He, R. J. Paull, J. D. Clarkson, J. Kudrnovsky, I. Turek, J. Kunes, D. Yi, J-H. Chu, C. T. Nelson, L. You, E. Arenholz, S. Salahuddin, J. Foncuberta, T. Jungwirth, and R. Ramesh, Nature Mater. \textbf{13}, 367 (2014).
	
\bibitem{Tveten:prl2014} E. G. Tveten, A. Qaiumzadeh, and A. Brataas, Phys. Rev. Lett. \textbf{112}, 147204 (2014).
	
\bibitem{Cheng:prl2014} R. Cheng, J. Xiao, Q. Niu, and A. Brataas, Phys. Rev. Lett. \textbf{113}, 057601 (2014).

\bibitem{Ohnuma:prb2013} Y. Ohnuma, H. Adachi, E. Saitoh, and S. Maekawa, Phys. Rev. B \textbf{87}, 014423 (2013).

\bibitem{Bauer:nmat2012} G. E. W. Bauer, E. Saitoh, and B. J. van Wees, Nature Mat. \textbf{11}, 391 (2012).

\bibitem{Uchida:nat2008} K. Uchida, S. Takahashi, K. Harii, J. Ieda, W. Koshibae, K. Ando, S. Maekawa, and E. Saitoh, Nature \textbf{455}, 778 (2008).


\bibitem{Xiao:prb2010} J. Xiao, G. E. W. Bauer, K. Uchida, E. Saitoh, and S. Maekawa, Phys. Rev. B \textbf{81}, 214418 (2010).

\bibitem{Adachi:prb2011} H. Adachi, J. Ohe, S. Takahashi, and S. Maekawa, Phys. Rev. B \textbf{83}, 094410 (2011).

\bibitem{Uchida:nmat2010} K. Uchida, J. Xiao, H. Adachi, J. Ohe, S. Takahashi, J. Ieda, T. Ota, Y. Kajiwara, H. Umezawa, H. Kawai, G. E. W. Bauer, S. Maekawa, and E. Saitoh,  Nature Materials \textbf{9}, 894 (2010).


\bibitem{Flipse:nntech2012} J. Flipse, F. L. Bakker, A. Slachter, F. K. Dejene, and B. J. van Wees \textbf{7}, 166-168 (2012).

\bibitem{Flipse:prl2014} J. Flipse, F. K. Dejene, D. Wagenaar, G. E. W. Bauer, J. Ben Youssef, B. J. van Wees, Phys. Rev. Lett. \textbf{113}, 027601 (2014).

\bibitem{Hoffman:prb2013} S. Hoffman, K. Sato, and Y. Tserkovnyak, Phys. Rev. B \textbf{88}, 064408 (2013).

\bibitem{Tserkovnyak:prl2002} Y. Tserkovnyak, A. Brataas, and G. E. W. Bauer, Phys. Rev. Lett. \textbf{88}, 117601 (2002).

\bibitem{Brataas:prb2002} A. Brataas, Y. Tserkovnyak, G. E. W. Bauer, and B. I. Halperin, Phys. Rev. B \textbf{66}, 060404 (2002).

\bibitem{Tserkovnyak:rmp2005} Y. Tserkovnyak, A. Bratas, G. E. W. Bauer, and B. I. Halperin, Rev. Mod. Phys. \textbf{95}, 016601 (2005).

\bibitem{Mizukami:jjap2001} S. Mizukami, Y. Ando, and T. Miyazaki, Jpn. J. Appl. Phys. \textbf{40}, 580 (2001).

\bibitem{Urban:prl2001} R. Urban, G. Woltersdorf, and B. Heinrich, Phys. Rev. Lett. \textbf{87}, 217204 (2001).


\bibitem{Hatami:prl2007} M. Hatami, G. E. W. Bauer, Q. Zhand, and P. J. Kelly, Phys. Rev. Lett. \textbf{99}, 066603 (2007).

\bibitem{Foros:prl2005} J. Foros, A. Brataas, Y. Tserkovnyak, and G. E. W. Bauer, Phys. Rev. Lett. \textbf{95}, 016601 (2005).

\bibitem{Kapelrud:prl2013} A. Kapelrud and A. Brataas, Phys. Rev. Lett. \textbf{111}, 097602 (2013).

\bibitem{Callen:prb1951} H. B. Callen and T. A. Welton, Phys. Rev. \textbf{83}, 34 (1951).

\end{thebibliography}
\end{document}